\documentclass[letterpaper, 10 pt, conference]{ieeeconf}  

\IEEEoverridecommandlockouts                              

\usepackage{amsmath} 
\usepackage{amssymb}  

\usepackage{amsthm}

\usepackage{gould_macros}
\usepackage{listings}
\usepackage{xcolor, enumerate}
\usepackage{graphicx, svg}
\usepackage[noEnd=true]{algpseudocodex}
\usepackage{algorithm}
\usepackage{caption, subcaption}
\usepackage{hyperref}

\hypersetup{
    colorlinks,
    linkcolor={red!50!black},
    citecolor={blue!50!black},
    urlcolor={blue!80!black}
}

\DeclareCaptionFormat{custom}
{%
  {\footnotesize#1#2 #3}
}
\theoremstyle{definition}

\newtheorem{example}{Example}

\newtheorem{problem}{Problem}

\newcommand{\meq}{m_{\rm eq}}
\newcommand{\Aeq}{A_{\rm eq}}
\newcommand{\beq}{b_{\rm eq}}
\newcommand{\mub}{m_{\rm ub}}
\newcommand{\Aub}{A_{\rm ub}}
\newcommand{\bub}{b_{\rm ub}}
\newcommand{\ncan}{n}
\newcommand{\mcan}{m}

\newcommand{\tableau}{\calT}
\newcommand{\cin}{c_{\rm enter}}
\newcommand{\rout}{r_{\rm exit}}

\newcommand{\kw}[1]{\lstinline!#1!}
\newcommand{\scipy}{\lstinline!scipy!}
\newcommand{\gurobi}{\lstinline!gurobi!}
\newcommand{\cvxpy}{\lstinline!CVXPY!}
\newcommand{\jaxopt}{\lstinline!JAXopt!}
\newcommand{\immrax}{\lstinline!immrax!}
\newcommand{\linrax}{\lstinline!linrax!}
\newcommand{\mpax}{\lstinline!MPAX!}

\makeatletter
\newcommand{\removelatexerror}{\let\@latex@error\@gobble}
\makeatother

\title{\LARGE \bf
\linrax{}: A JAX Compatible, Simplex Method Linear Program Solver
}

\author{Brendan Gould, Akash Harapanahalli, and Samuel Coogan$^{1}$
\thanks{*This work was supported in part by the National Science Foundation under awards \#2219755 and \#2333488 and by the Air Force Office of Scientific Research under Grant FA9550-23-1-0303.}
\thanks{$^{1}$The authors are affiliated with the School of Electrical and Computer Engineering, Georgia Institute of Technology,
        Atlanta, GA 30332, USA
        {\tt\small \{bgould6, aharapan, sam.coogan\}@gatech.edu}.%
}}

\newcommand{\ie}{\emph{i.e.}}
\newcommand{\eg}{\emph{e.g.}}

\newcommand{\conc}[2]{\begin{bmatrix} #1 \\ #2 \end{bmatrix}}

\newcommand{\calH}{\mathcal{H}}

\newcommand{\calO}{\mathcal{O}}

\newcommand{\calR}{\mathcal{R}}

\newcommand{\calT}{\mathcal{T}}

\newcommand{\calW}{\mathcal{W}}
\newcommand{\calX}{\mathcal{X}}

\newcommand{\bfw}{\mathbf{w}}

\newcommand{\bfI}{\mathbf{I}}

\newcommand{\ul}[1]{\underline{#1}}

\newcommand{\ulw}{\ul{w}}
\newcommand{\ulx}{\ul{x}}
\newcommand{\uly}{\ul{y}}
\newcommand{\ulz}{\ul{z}}

\newcommand{\ol}[1]{\overline{#1}}

\newcommand{\olw}{\ol{w}}
\newcommand{\olx}{\ol{x}}
\newcommand{\oly}{\ol{y}}
\newcommand{\olz}{\ol{z}}

\begin{document}
 \lstset{
    language=Python,
    basicstyle=\ttfamily
}

\maketitle
\thispagestyle{empty}
\pagestyle{empty}

\begin{abstract}

We present \linrax{}, the first simplex based linear program (LP) solver in the JAX ecosystem.
In many control algorithms, 
LPs are often automatically generated and solved online in the control loop. 
This motivates the design of \linrax{}, which is especially suited for compilation into a large JAX-based pipeline as a subroutine.
We discuss the challenges of implementing a general purpose LP solver under strict design requirements from JAX.
Notably, we can solve general problems which may include dependent constraints---something not possible with existing JAX-compatible LP solvers that use first-order techniques and may fail to converge.
We demonstrate \linrax{}'s utility through several examples, including a robust control synthesis pipeline for nonlinear vehicles using automatic differentiation through a LP-based reachable set framework.

\end{abstract}

\section{Introduction}
\label{sec:intro}

Constrained optimization is an important tool in robotics and control and appears in many forms.  
For example, techniques like Model-Predictive Control (MPC)~\cite{anderssonCasADiSoftwareFramework2019}, Control-Lyapunov functions (CLFs)~\cite{sontagControlLyapunovFunctions1999}, and Control-Barrier functions (CBFs)~\cite{amesControlBarrierFunctions2019} 
choose control inputs \emph{at each point in time} as the solution of an optimization problem. 
Linear programs (LPs)~\cite{dantzigLinearProgrammingExtensions2016} are a valuable subset of optimization problems that are capable of synthesizing plans or control policies~\cite{bahreinianRobustPathPlanning2021} and verifying the safe operation of control systems through reachability analysis~\cite{harwoodEfficientPolyhedralEnclosures2016}. 
Optimization-based control frameworks have been facilitated by modern advances in computational power, including a rich ecosystem of software optimizers~\cite{2020SciPy-NMeth, gurobi}.
Many such solvers support additional features such as automatic differentiation~\cite{diamond2016cvxpy} and GPU parallelization~\cite{spampinato2009linear}, making it possible to embed optimization problems into gradient-based, scalable control pipelines. 

JAX is a high-performance computing platform for Python that provides Just-in-Time Compilation (JIT), GPU Parallelization, and automatic differentiation~\cite{jax2018github}, increasingly being used for control applications~\cite{harapanahalliImmraxParallelizableDifferentiable2024, grillottiKheperaxLightweightJAXbased2023, rutherfordJaxMARLMultiAgentRL2024}. 
Though there exist many JAX libraries for optimization~\cite{deepmind2020jax, optimistix2024} or gradient descent~\cite{kidger2021equinox}, surprisingly few are designed for mathematical programming problems. 
Those that are~\cite{jaxopt_implicit_diff, lu2024mpax} utilize first-order gradient based methods and struggle with solution precision and convergence, especially on problems with degenerate constraints. 
To address this gap, we introduce a JAX implementation of the \emph{simplex algorithm}, a method for solving LPs exactly (up to floating-point accuracy) that works for problems with degenerate constraints, and demonstrate its utility in control algorithms. 

\subsection{Contributions}

This work presents \linrax{}, an open-source implementation of the simplex algorithm in JAX, available on PyPI and GitHub\footnote{The PyPI package name is \linrax{}, and the repository url is \kw{https://github.com/gtfactslab/linrax}.}. 
Our implementation modifies the classic algorithm to satisfy strict JAX \emph{tracibility} requirements, and therefore enjoys JIT compilation, automatic differentiability, and GPU parallelization. 
It is especially suited for solving relatively small-sized LPs with high precision, and when embedded as a subroutine in more complex JAX programs (as is common in modern control pipelines).
Compared to other JAX optimizers, \linrax{} easily handles problems with linearly dependent constraints, something which causes convergence delays or even outright failures in gradient-based methods. 

In Section~\ref{sec:safe_control}, we use \linrax{} to solve a safe control problem by computing a control ``nudge'' that minimally perturbs a nominal input trajectory to ensure robust safety, leveraging reachability analysis and automatic differentiation through LPs. 
Section~\ref{sec:lps_in_JAX} pinpoints the incompatibility between JAX and traditional simplex-based approaches, namely the handling of dependent constraints. 
Our solution, special handling of the auxiliary variables in the ratio test, is described in Section~\ref{subsec:jax_tracing}.
Finally, Section~\ref{sec:contributions} examines the interface of \linrax{}, and compares it to existing LP solvers, both in and outside of the JAX ecosystem. 
We show that \linrax{} is competitive with finely-tuned solvers on relatively small problems, and when embedded in a more complex optimization-based pipeline as intended, \linrax{} excels compared to these other solvers. 


\subsection{Notation}

Let $\ol\R = \R\cup\{\pm\infty\}$.
For $x,y\in\ol\R^n$, we use the elementwise partial order, \ie, $x\leq y$ when $x_i\leq y_i$ for every $i=1,\dots,n$. 
For $\ulx\leq\olx$, define the interval $[\ulx,\olx] = \{x\in\ol\R^n : \ulx \leq x \leq \olx\}$.
$\bfI_n$ is the $n\times n$ identity matrix, and $0_{m\times n}$ ($1_{m \times n}$) is a $m\times n$ matrix of zeros (ones).
We use brackets to index arrays: for a matrix $A$ and $i, j \in \Z^+$, $A[i, j]$ is the component of $A$ in the $i$th row and $j$th column, $A[i, :]$ is the entire $i$th row of $A$, and $A[\text{end}, \text{end}]$ is the entry in the last row and column of $A$. 
Given $x,y\in\R^n$, $x_{i:y}\in\R^n$ is the vector $(x_{i:y})_j = x_j$ for every $j\neq i$ and $(x_{i:y})_i = y_i$.

\section{Control Nudging for Robust Safety}

\label{sec:safe_control}

LPs are a staple in many safety enforcing control algorithms. 
Here, we use them to modify a nominal control input into a guaranteed safe input using reachability analysis and automatic differentiation through an LP.
This section serves as both motivation for \linrax{} and our main case study.

Given a nonlinear control system of the form
\begin{align} \label{eq:nlsys}
    \dot{x} = f(x,u,w),
\end{align}
where $x\in\R^n$ is the state, $u\in\R^p$ is the control, $w\in\R^q$ is a disturbance, and $f:\R^n\times\R^p\times\R^q$ is a locally Lipschitz vector field, reachability analysis aims to compute the set of states the system can reach for initial conditions $\calX_0\subset\R^n$, a feed-forward control mapping $u_\mathrm{ff}:\R\to\R^p$, and disturbances $\calW\subset\R^p$.
Formally, the reachable set of~\eqref{eq:nlsys} at time $t$ is 
\begin{align*}
    \calR(t,\calX_0,u_\mathrm{ff},\calW) = \{\phi^f(t,x_0,u_\mathrm{ff},\bfw),\,\bfw:\R\to\calW\},
\end{align*}
where $\phi^f$ is the flow of~\eqref{eq:nlsys} from initial condition $x_0 \in \calX_0$ under inputs $u_\mathrm{ff}(t)$ and $\bfw(t)$ at time $t$.
If the full reachable set is safe, \eg, avoids an obstacle $\calO\subseteq\R^n$, then the system is robust to initial condition and/or applied disturbances.

\begin{problem}[Feedforward control filtering]
We seek to minimally perturb a nominal feedforward input $u_\mathrm{ff}:[0,T]\to\R^p$ such that the reachable set avoids $\calO$, \ie, find $u'_\mathrm{ff}$ such that
\begin{align*}
    \calR(t,\calX_0,u'_\mathrm{ff},\calW) \cap \calO = \emptyset, \quad \forall \, t \in [0, T].
\end{align*}
\end{problem}

\begin{example}
\label{ex:bicycle}
Consider the dynamics of a kinematic bicycle,
\begin{align}
\label{eq:kinematic_bicycle}
\begin{aligned}
    \dot{p}_x &= v \cos(\phi + \beta(u_2)), & \dot{\phi} &=\frac{v}{\ell_r}\sin(\beta(u_2)), \\
    \dot{p}_y &= v \sin(\phi + \beta(u_2)), & \dot{v} &= u_1,
\end{aligned}
\end{align}
where $[p_x\ p_y]^\top\in\R^2$ is the $x$-$y$ displacement of the center of mass (COM), $v\in \R_{\geq 0}$ is its speed, and $\phi \in [-\pi,\pi)$ is the heading angle.
Input $u_1$ is the applied force, $u_2$ is the angle of the front wheel, and $\beta(u_2) = \mathrm{arctan}\left(\frac{\ell_f}{\ell_f+\ell_r}\tan(u_2)\right)$ is the slip slide angle where $\ell_f$ ($\ell_r$) is the distance between the COM and the front (rear) wheel.
For simplicity, we use $\ell_f = \ell_r = 1$, and Euler integrate with $\Delta t = 5\cdot 10^{-3}$.

Let $o(x) = 3^2 - (p_x - 4)^2 - (p_y - 4)^2$, $\calO = \{x\in\R^4 : o(x) \geq 0\}$ denote a circular obstacle.
We first generate a nominal control strategy $u_\mathrm{ff}$ using nonlinear MPC in \texttt{casadi}~\cite{anderssonCasADiSoftwareFramework2019}, to stabilize the vehicle to the origin while avoiding the obstacle.
This is done using LQR with $A = \frac{\partial f}{\partial x}(\mathring{x},0,0)$, $B = \frac{\partial f}{\partial u}(\mathring{x},0,0)$ linearized about the initial state $\mathring{x} = \begin{bmatrix} 8 & 7 & -\frac{2}{\pi} & 2 \end{bmatrix}$ to obtain a feedback matrix $K$ with $Q = \bfI_4$, $R = \bfI_2$.
Given the feedforward $u_\mathrm{ff}$ and the feedback matrix $K$, we analyze the safety of the closed-loop system
\begin{align}
    \label{eq:cl_bicycle}
    \dot{x} = f(x, u_\mathrm{ff}(t) + K(x(t) - x_\mathrm{nom}(t)), w).
\end{align}
\end{example}

The MPC trajectory tightly follows the obstacle, meaning disturbances may cause a collision (Figure~\ref{fig:vehicle_reachsets}, left) .
Using \linrax{} and reachability analysis, we modify $u_\mathrm{ff}$ so that trajectories of~\eqref{eq:cl_bicycle} robustly avoid the obstacle.





\subsection{Linear Programs for Reachable Set Overapproximation}
\label{subsec:LP_reachset}

In practice, the true reachable set is difficult to obtain and approaches instead verify the safety of an \emph{overapproximation} $\ol\calR(t,\calX_0,\calW) \supseteq \calR(t,\calX_0,\calW)$. 
One method uses interval analysis~\cite{Jaulin:2001} to compute $\ol\calR$ (Figure~\ref{fig:vehicle_reachsets}, left). 
Later work~\cite{shenRapidAccurateReachability2017, harwoodEfficientPolyhedralEnclosures2016} uses \emph{interval refinement} via LPs to produce a less conservative, polytopic overapproximation. 
We briefly recall this approach here, but refer to the original works for details. 

\begin{figure}
\vspace{1mm}
    \centering
    \includesvg[width=\columnwidth]{fig/reachable_sets_combined_20_22_24.svg}
    \caption{%
        Overapproximations of the reachable sets of a kinematic bicycle. 
        Using basic interval analysis, the bounds are too conservative for use in safety filtering (\textbf{left}). 
        With LP refinement~\eqref{eq:LP_refine}, the bounds are tighter, but the feedforward control strategy is unsafe and a sample unsafe trajectory is shown in green (\textbf{center}). 
        Algorithm~\ref{alg:control_filtering} modifies the feedforward control strategy so that the entire reachable set avoids the obstacle (\textbf{right}).
    }
    \label{fig:vehicle_reachsets}
    \vspace{-1mm}
\end{figure}



Trajectories of~\eqref{eq:nlsys} sometimes satisfy an invariance condition: $\phi^f(t,x_0,u_\mathrm{ff},\bfw) \in \calH$, arising from some conservation law, artificially introduced by adding model redundancies~\cite{shenRapidAccurateReachability2017}, or by lifting the system to higher dimensions~\cite{harapanahalliCertifiedRobustInvariant2024}.
This property can be combined with any bound $[\uly, \oly]$ from interval analysis to significantly reduce conservatism (especially if $\calH$ is well-chosen).
Any interval $[\ulz, \olz]$ with
\begin{align} \label{eq:LP_refine}
    \calH\cap[\uly,\oly]\subseteq [\ulz,\olz] \subseteq[\uly,\oly],
\end{align}
is a \emph{refinement} of $[\uly, \oly]$ that still contains the system's reachable set. 
For $\calH = \{Hx : x\in\R^n\}$ (with $H\in\R^{m\times n}$ full rank), the tightest refinement is given by the LPs
\begin{subequations}
\begin{align}
    \ulz_j &= \min_{x\in\R^n} e_j^\top Hx \ \text{ s.t. } \ \uly\leq Hx \leq \oly, \\
    \olz_j &= \max_{x\in\R^n} e_j^\top Hx \ \text{ s.t. } \ \uly\leq Hx \leq \oly.
\end{align}
\end{subequations}
In~\cite{shenRapidAccurateReachability2017,harapanahalliCertifiedRobustInvariant2024}, LPs were avoided using a different refinement via dual space sampling, at the cost of looser bounds.
Instead, we use \linrax{} to embed these LPs directly in the reachability framework, while retaining JIT compilation and automatic differentiation.

\begin{figure}
\vspace{-2mm}
\begin{algorithm}[H]
\caption{Safe Control Filtering Using Automatic Differentiation of LPs}
\label{alg:control_filtering}
\begin{algorithmic}[1]
    \State \textbf{Input:} full rank $H\in\R^{m\times n}$, left inverse $H^+$ such that $H^+H = I_n$, initial set $\{x\in\R^n : \uly_0\leq Hx \leq \oly_0\}$, initial control map $u_\mathrm{ff}$, disturbance set $[\ulw,\olw]$, obstacle set $\{x\in\R^n : o(x) \geq 0\}$, time horizon $T$
    \Function{RefinedEmbedding}{$u_\mathrm{ff}$}
        \For {$i\in\{1,\dots,m\}$}
            \State $\ulz\gets\uly$, $\olz\gets\oly_{i:\uly}$
            \State $\ulz_j \gets \min_x e_j^\top Hx \ \text{ s.t. } \ \uly \leq Hx \leq \oly_{i:\uly}$, $\forall j$ \label{alg:control_filtering:flattening_1}
            \State $\olz_j \gets \max_x e_j^\top Hx \ \text{ s.t. } \ \uly \leq Hx \leq \oly_{i:\uly}$, $\forall j$ \label{alg:control_filtering:flattening_2}
            \State $\dot{\uly}_i \gets \min_{y\in[\ulz,\olz],w\in[\ulw,\olw]} f_i(H^+y,u_\mathrm{ff}(t),w)$ 
            \Statex \Comment{underapproximated using interval analysis}
            \State $\ulz\gets\uly_{i:\oly}$, $\olz\gets\oly$
            \State $\ulz_j \gets \min_x e_j^\top Hx \ \text{ s.t. } \ \uly_{i:\oly} \leq Hx \leq \oly$, $\forall j$ \label{alg:control_filtering:flattening_3}
            \State $\olz_j \gets \max_x e_j^\top Hx \ \text{ s.t. } \ \uly_{i:\oly} \leq Hx \leq \oly$, $\forall j$ \label{alg:control_filtering:flattening_4}
            \State $\dot{\oly}_i \gets \max_{y\in[\ulz,\olz],w\in[\ulw,\olw]} f_i(H^+y,u_\mathrm{ff}(t),w)$ \hfill
            \Statex \Comment{overapproximated using interval analysis}
        \EndFor
    \EndFunction
    \Function{SafetyCheck}{$u_{\mathrm{ff}}$}
        \State $\{[\uly(t),\oly(t)]\}_{t\in[0,T]} \gets$ \textproc{RefinedEmbedding} System Trajectory
        \State $\ol\calR_t \gets \{x : \uly(t) \leq Hx \leq \oly(t)\}$, for every $t\in[0,T]$
        \State \Return $\int_0^T \max\{\sup_{x\in\ol\calR_t} g(x),0\} \ \mathrm{d}t$
    \EndFunction
    \Repeat
        \State $u_{\mathrm{ff}} \gets u_{\mathrm{ff}} - \eta\nabla_{u_\mathrm{ff}}\textproc{SafetyCheck}(u_\mathrm{ff})$  \label{alg:control_filtering:grad}
    \Until{$\textproc{SafetyCheck}(u_\mathrm{ff}) \leq 0$}
    \State \Return $u_\mathrm{ff}$
\end{algorithmic}
\end{algorithm}
\vspace{-8mm}
\end{figure}

In Algorithm~\ref{alg:control_filtering}, interval analysis and refinement as in~\eqref{eq:LP_refine} build the \textproc{RefinedEmbedding} system, whose trajectory $[\uly(t),\oly(t)]$ overapproximates the reachable set as a polytope,
\begin{align*}
    \calR(t,\{x\in\R^n &: \uly\leq Hx\leq \oly\},[\ulw,\olw]) \\
    & \subseteq\{x\in\R^n : \uly(t) \leq Hx \leq \oly(t)\}.
\end{align*}
Note that the component replacements on Lines~\ref{alg:control_filtering:flattening_1}-\ref{alg:control_filtering:flattening_2} and \ref{alg:control_filtering:flattening_3}-\ref{alg:control_filtering:flattening_4} \emph{guarantee} the inclusion of degenerate constraints in the generated LP, inhibiting the usage of existing JAX solvers. 

\subsection{Safety Filtering Using Automatic Differentiation}

We introduce invariance to~\eqref{eq:kinematic_bicycle} through the \emph{lifting} matrices
\begin{align*}
    H = \conc{\bfI_n}{H_2}, \quad 
    H_2 = \left[\begin{smallmatrix}
        1 & 0 & 1 & 0 \\ 
        1 & 0 & -1 & 0 \\
        0 & 1 & 1 & 0 \\
        0 & 1 & -1 & 0
    \end{smallmatrix} \right], \quad
    H^+ = [\bfI_n\ 0_{n\times n}],
\end{align*}
and apply Algorithm~\ref{alg:control_filtering} to generate refined reachable sets (Figure~\ref{fig:vehicle_reachsets}, center).
Informally, these matrices ``couple'' the position states with the heading angle, allowing refinement to exploit their connection through the dynamics. 
This tightens the bounds, allowing a small ``nudge'' to separate them from the obstacle. 
Thanks to JAX and \linrax{}'s automatic differentiation, computing this nudge is straightforward. 
We quantify the size of $\ol \calR_t \cap \calO$: $\max\left\{\sup_{x\in\ol\calR_t} o(x),0\right\}$, then sum over time to obtain a function $\textproc{SafetyCheck}$ mapping a feedforward input to a number that is positive if the reachable set intersects the obstacle and $0$ otherwise. 
Finally, we apply gradient descent in $u_\mathrm{ff}$ to modify the input until $\textproc{SafetyCheck}(u_\mathrm{ff}) = 0$, giving a robustly safe trajectory (Figure~\ref{fig:vehicle_reachsets}, right).
This example took $8$ gradient descent steps in $195$s (after $52$s of JIT compilation).
For now, \linrax{} only supports forward-mode autodifferentiation, which contributes to the computation time.
Since \textproc{SafetyCheck} has many inputs (the nominal control inputs at the sampled time steps) and one output (the safety value), reverse-mode autodifferentiation may be better suited. 
Though we use Example~\ref{ex:bicycle} as a motivating case study, we emphasize that the core computational capabilities of \linrax{}---JIT compilation, GPU parallelization, automatic differentiation, and the ability to solve problems with dependent constraints---have a broad scope of control applications. 

\section{Linear Programs in JAX}
\label{sec:lps_in_JAX}

We now describe the general form of LPs and explain the difficulty associated with solving them in JAX. 
In particular, linearly dependent constraints cannot be handled in the same way as other solvers. 
We present a modification of the tableau simplex method to address this limitation.

\subsection{Linear Programs}

An LP is an optimization problem with a linear objective and linear (equality or inequality) constraints:
\begin{equation}
    \min_{x \in [l, u]} c^\top x \text{ s.t. } \begin{array}{l} \Aeq x = \beq, \\ \Aub x \le \bub, \end{array}
    \label{eq:LP_Def_General}
\end{equation}
where $x\in\R^n$, $l, u \in \ol{\R}^n$, $l\leq u$, $\Aeq \in \R^{\meq \times n}$, $\beq \in \R^{\meq}$, $\Aub \in \R^{\mub \times n}$, $\bub \in \R^{\mub}$, and $c \in \R^n$. 
Most computational solvers specify objective, equality constraints, inequality constraints, and variable bounds separately, as in~\eqref{eq:LP_Def_General}. 
However, in theory, it suffices to consider a simplification: any such problem can be written in \emph{canonical form}
\begin{equation}
    \min_{x \ge 0_{n}} c^\top x \text{ s.t. } Ax = b, \quad b \ge 0_m,
    \label{eq:LP_Def_Canonical}
\end{equation}
(with larger dimension $A \in \R^{\mcan \times \ncan}$, $b\in \R^\mcan$, $c \in \R^\mcan$). 
Inequality constraints are enacted by adding new, non-negative ``slack'' variables. 
To minimize over alternative domains, express each input as the difference between two positive components and add constraints as needed. 

Converting a problem to canonical form is called \emph{pre-solving} and may also determine the problem's feasibility or boundedness~\cite{andersenPresolvingLinearProgramming1995} or eliminate redundant constraints~\cite{andersenFindingAllLinearly1995}. 
Pruning dependent constraints not only reduces the problem's dimension, it also ensures that the matrix $A$ is full rank, a necessary condition for many solvers~\cite{huangfuParallelizingDualRevised2018, pmlr-v37-nishihara15}. 
Unfortunately, this simplification is incompatible with the JAX ecosystem, and will be examined further in Section~\ref{subsec:jax_tracing}.

\subsection{The Simplex Method}
\label{subsec:simplex_method}

One common LP solver is the \emph{simplex method}, which we briefly recall here to fix notation (for more details, see~\cite{dantzigLinearProgrammingExtensions2016}). 
This method iteratively travels between vertices of the feasible polytope, improving the objective at each iteration. 
When no adjacent vertex is an improvement, an optimal solution has been found and the algorithm terminates. 
It often runs in two phases; the first phase identifying a vertex of the feasible region which initializes the second phase to solve the original problem. 
The first phase solves the \emph{auxiliary problem}
\begin{equation}
   \min_{x, a \ge 0} \begin{bmatrix} 0_{1 \times n} & 1_{1 \times m} \end{bmatrix} \begin{bmatrix} x \\ a\end{bmatrix} \text{ s.t. } \begin{bmatrix} A & \bfI_{\mcan} \end{bmatrix} \begin{bmatrix} x \\ a \end{bmatrix} = b. 
   \label{eq:simplex_phase1}
\end{equation}
Note that $x = 0_{1 \times \ncan}$, $a = b$ is an initial feasible vertex of~\eqref{eq:simplex_phase1}. 
For any optimal $(x^*, a^*)$, if $a^* = 0_{1 \times \mcan}$, then $x^*$ is a feasible vertex of~\eqref{eq:LP_Def_Canonical}, as desired. 
Otherwise,~\eqref{eq:LP_Def_Canonical} is infeasible. 

The state of the simplex method is maintained in a \emph{tableau}. 
In the first phase, this tableau is initialized to 
\begin{equation}
\label{eq:tableau}
    \tableau_1 = \begin{bmatrix}
        A & \bfI_{\mcan} & b \\
        c & 0_{1 \times \mcan} & -z \\ 
        -\sum_{i=1}^{m} A[i, :] & 0_{1 \times \mcan} & -w
    \end{bmatrix}.
\end{equation}
The upper block represents the constraints, and the bottom two rows give the \emph{reduced cost multipliers} of the original and auxiliary problem. 
The current feasible vertex is parameterized by a set of \emph{basic variables}; here, those at indices $\ncan + 1, \ldots, \ncan + \mcan$.
This structure is maintained throughout every iteration by choosing one variable to enter the set of basic variables and forcing another to exit. 
The \emph{pivoting strategy} that selects the entering variable is key.
In \linrax{}, we use \emph{Bland's Rule}~\cite{avisNotesBlandsPivoting1978}, a simple approach guaranteed to never cycle. 
If any index has negative cost multiplier, it may improve the objective when entering the active set; thus, optimization continues until no such index remains. 

After choosing an entering variable ($\cin$), the exiting variable is deduced by the \emph{ratio test}. 
Using Gauss-Jordan elimination, we cause the column of $x[\cin]$ to become basic, giving each current basic variable a value of $x[r] = b[r] - A[r, \cin] x[\cin]$. 
To maintain feasibility, $x[\cin]$ cannot increase past $\frac{b[r]}{A[r, \cin]}$ for any $r$ where $A[r, \cin] > 0$.
The row which minimizes this ratio is the exiting variable. 
If no such row exists, then $x[\cin]$ (and~\eqref{eq:LP_Def_Canonical}) are unbounded. 

After phase 1 is solved, the resulting tableau initializes phase 2. 
By discarding columns $\ncan +1, \ldots, \ncan + \mcan$ (for the auxiliary variables) and the bottom row (auxiliary cost), a tableau for the original problem is produced.
If the problem is \emph{degenerate} (\emph{e.g.}, through linearly dependent constraints), auxiliary variables may remain in the active set, and these variables' rows can also be discarded. 
(Though theoretically valid, we shall see this is impossible to implement in JAX.)
Finally, the pivoting procedure is repeated again until no further improvement is possible, yielding an optimal point. 

\subsection{JAX and \linrax{}}
\label{subsec:jax_tracing}

We now describe \linrax{}, a JAX implementation of the simplex method. 
JAX is a library providing \emph{function transformations} that map one function to another; \emph{e.g.}
\kw{jacfwd} / \kw{jacrev} return the input's derivative using forward- or reverse-mode automatic differentiation, while \kw{jit} ``compiles'' the input into equivalent XLA operations. 
To compute these transformations, JAX first \emph{traces} the input and represents its computation graph as a \kw{jaxpr}. 
JAX cannot trace arbitrary Python and restricts the operations allowed in transformable functions~\cite{jax2018github}. 
In particular, all arrays must have static shape (\emph{i.e.}, deducible at JIT-compile time). 
This forbids modifications to the static \emph{shape} of an array based on its traced \emph{values}---a problem for the simplex method. 

While the constraint matrix $A$ has static shape, the number of redundant constraints is dependent on its values, so such constraints cannot be pruned in a pre-solve step as is common~\cite{2020SciPy-NMeth}, nor can we discard those remaining in the active set after phase 1.
Non-traceable code that removes dependent constraints can be written, but only to solve LPs in isolation. 
If solving a LP is merely one part of a pipeline, as is often the case for controls applications (see Section~\ref{sec:safe_control}), then the entire pipeline would no longer benefit from JAX transformations. 


To overcome this incompatibility, we introduce Algorithm~\ref{alg:simplex}, a modification of the simplex method. 
We largely follow standard practice, but modify the ratio test during the second phase. 
Since we do not assume full rank $A$, auxiliary variables may remain in the active set after phase one, and JAX prevents us from discarding these variables as is standard. 
Instead, during the ratio test in phase 2, we use the \emph{absolute} rate of change for these variables rather than ignoring negative components. 
In feasible problems, every auxiliary variable is 0 after phase 1. 
Therefore, Algorithm~\ref{alg:simplex}, Line~\ref{alg:simplex:pivot_gate} ensures that whenever an entering variable that would affect the value of any remaining auxiliary variables is chosen, that variable gets 0 in the ratio test\footnote{A previous version of this work used a different approach that terminated with a sub-optimal solution on some problems.}. 
Since $b \ge 0$ by~\eqref{eq:LP_Def_Canonical}, this is the lowest ratio possible, and these variables exit the active set. 
Once they exit, they will not enter again, since~\eqref{eq:LP_Def_General} does not depend on auxiliary variables.

\begin{figure}
\vspace{-2mm}
\begin{algorithm}[H]
\caption{The Simplex Method in JAX}
\label{alg:simplex}
\begin{algorithmic}[1]
    \Require Pivot strategy \textproc{SelectEnteringVar}
    \State \textbf{Input:} Constraint matrix $A \in \R^{m \times n}$, constraint vector $0 \le b \in \R^m$, objective vector $c \in \R^n$ of the form~\eqref{eq:LP_Def_Canonical}. 
    \Function{Pivot}{tableau $\tableau$, basic indices $I$}
        \State $\cin \gets$ \textproc{SelectEnteringVar}$(\tableau)$
        \State $L \gets \{r : \tableau[r, \cin] > 0 \lor I[r] > n\}$ \Comment{Moving toward boundary OR remaining auxiliary variable} \label{alg:simplex:pivot_gate}
        \If{$L = \emptyset$}
            \State \Return The problem is unbounded
        \EndIf
        \State $\rout \gets \arg\min_{r \in L} \tableau[r, -1] / \lvert \tableau[r, \cin] \rvert$
        \State $I[\rout] \gets \cin$
        \State $\tableau[\rout, :] \gets \tableau[\rout, :] / \tableau[\rout, \cin]$
        \For{$r \in \{ 1, 2, \ldots, \text{end} \} \setminus \{\rout\}$}
        \State $\tableau[r, :] \gets \tableau[r, :] - \tableau[r, \cin] \cdot \tableau[\rout, :]$ 
        \EndFor
        \State \Return $\tableau$, $I$ 
        
    \EndFunction

    \State $I \gets \{n+1, n+2, \ldots, n+m\}$ 
    \State $\tableau_1 \gets$ form of~\eqref{eq:tableau}
    \While{Any $\tableau_1[\text{end}, 1:\text{end}-1] < 0$} 
        \State $\tableau_1, I \gets$ \textproc{Pivot}($\tableau_1$, $I$)
    \EndWhile
    \If{$c_{aux} \cdot x_{aux} > 0$}
        \State \Return The problem is infeasible
    \EndIf

    \State $\tableau_2 \gets \texttt{hstack}(\tableau_1[:\text{end}, :n+1], \tableau_1[:\text{end}, \text{end}])$
    \While{Any $\tableau_2[\text{end}, :\text{end}] < 0$} 
        \State \textproc{Pivot}($\tableau_2$, $I$)
    \EndWhile
    \State \Return $\tableau_2$ 
\end{algorithmic}
\end{algorithm}
\vspace{-8mm}
\end{figure}

\section{Interface \& Comparison}
\label{sec:contributions}

We describe the interface and usage of \linrax{} in~\ref{subsec:interface}, presenting the main function's signature.
In~\ref{subsec:comparison}, we compare \linrax{} to other LP solvers, including another JAX implementation. 
On smaller problems ($<50$ inputs), its performance is comparable to industry standard toolkits.
Additionally, \linrax{} is a fully composable JAX subroutine and can be used in large pipelines without sacrificing these advantages, which is not possible with most other solvers. 

\vspace{-1mm}
\subsection{Interface}
\label{subsec:interface}


The signature of the main function in \linrax{} is: 

{
\footnotesize
\begin{lstlisting}
@partial(jax.jit, static_argnames=["unbounded"])
def linprog(c: jax.Array, 
    A_ub: jax.Array, b_ub: jax.Array, 
    A_eq: jax.Array, b_eq: jax.Array,
    unbounded: bool
) -> Tuple[SimplexStep,  SimplexSolutionType]:
    ...
\end{lstlisting}
}

Here, the first five parameters are as in~\eqref{eq:LP_Def_General} and closely mimic \scipy{}'s interface. 
Importantly, we do not assume that \kw{A_ub} or \kw{A_eq} are full rank. 
The last parameter, \kw{unbounded}, determines if minimization is performed on positive inputs or an unbounded domain. 
Since this change internally doubles the number of input variables, \kw{unbounded} is a \emph{static argument}, so JAX treats it as a constant and re-traces the function whenever it is assigned a different value. 
This sidesteps the limitation from~\ref{subsec:jax_tracing} and allows its value to affect arrays' shapes. 
Since booleans have only two values, this is a manageable drawback. 

The \kw{SimplexSolutionType} object describes the LP.
This object contains three \kw{bool} fields: \kw{feasible}, \kw{bounded}, and \kw{success}, indicating the corresponding properties of the problem. 
If minimization was a \kw{success}, the \kw{SimplexStep} object is a solution. 
The optimal input and objective value are in fields \kw{x} and \kw{fun}, respectively. 
Furthermore, \kw{tableau} retrieves the full internal tableau, and \kw{basis} gives the indices of the basic variables at \kw{x}. 

\subsection{Comparison}
\label{subsec:comparison}

We examine \linrax{}'s performance relative to several off-the-shelf LP solvers. 
Both \scipy{}~\cite{2020SciPy-NMeth} and \gurobi{}~\cite{gurobi} implement specialized algorithms for LPs. 
These libraries only provide numerical solutions, and do not support any additional features. 
\cvxpy{}~\cite{diamond2016cvxpy} solves constrained, convex optimization problems, including LPs, and supports automatic differentiation through the optimization. 
A library~\cite{agrawal2019differentiable} integrates this gradient information with JAX, but is not composable with other transformations such as \kw{jit} or \kw{vmap}. 
There also exist first-order, JAX-compatible optimizers capable of solving LPs: \jaxopt{}~\cite{jaxopt_implicit_diff} and \mpax{}~\cite{lu2024mpax}\footnote{Due to the similarity of \jaxopt{}'s solver to that of \mpax{}, and the earlier stage of implementation of the latter, we compare only to the former.}.

To our knowledge, \linrax{} is the first JAX implementation of the simplex method, making it uniquely suited for use as a subroutine in large control pipelines. 
JAX compatibility provides JIT compilation, vectorization, and GPU parallelization, enabling efficient solving of many LPs simultaneously. 
Our simplex based algorithm provides an exact solution (up to numerical precision) and works on problems with dependent constraints. 
This is a clear advantage over existing gradient-based JAX optimizers, which converge to a tolerance (as high as $10^{-4}$~\cite{lu2024mpax}) and may fail entirely for such problems---even small ones. 
If the goal is to solve a single LP as fast as possible, other optimizers outperform \linrax{} through the use of specialized algorithms, though we are competitive on small problems (Table~\ref{table:small_comp}).
In the more practical case of solving LPs inside a larger computation, \linrax{}'s unique benefits become clear (Table~\ref{table:intref_comp}). 



All tests were run on a Kubuntu 22.04.05 system, with an Intel Xeon Gold 6320 CPU, Quadro RTX 8000 GPU, and 64GB of RAM. 
For sample size $N > 0$, each solver optimizes the same problem $N$ times, and we report the mean and standard deviation runtime. 

\begin{example}[Random, small linear program]
\label{ex:random_lp}
    A random LP  with $n=20$ inputs and $m_{ub} = 15$ inequality constraints is generated and optimized by each solver.
\end{example}
The results of this test are given in Table~\ref{table:small_comp}: \linrax{} exhibits performance on the same order of magnitude as professional solvers, beating the other JAX implementation. 

\begin{table}
\centering
\vspace{2mm}
\begin{tabular}{|c|c|c|}
    \hline
    Method & \kw{jit} (sec) & Time (sec) \\
    \hline \hline 
    \scipy{} & - & $\bf 1.082 \cdot 10^{-3} \pm 7.964 \cdot 10^{-4}$  \\ 
    \hline
    \gurobi{} & - & $2.23 \cdot 10^{-3} \pm 4.613 \cdot 10^{-3}$  \\ 
    \hline
    \cvxpy{} & - & $1.179 \cdot 10^{-3} \pm 8.879 \cdot 10^{-4}$  \\ 
    \hline
    \jaxopt{} &\bf{0.951} & $3.477 \cdot 10^{-2} \pm 3.497 \cdot 10^{-3}$ \\
    \hline
    \bf{\linrax{}} & 1.226 & $6.255 \cdot 10^{-3} \pm 7.874 \cdot 10^{-4}$ \\
    \hline
\end{tabular}
\caption{%
    Comparison of LP solvers on a small, random LP as described in Example~\ref{ex:random_lp}
    Sample size $N=100$.
}
\label{table:small_comp}
\end{table}

In the second test, we use each solver to compute an over-approximation of a dynamical system's reachable set.
\begin{example}[Reachable Set Approximation]
\label{ex:reach}
Consider the Van der Pol oscillator, described by the ODE 
\begin{equation}
\label{eq:vanderpol}
\begin{aligned}
    \dot{x}_1 = \mu \left(x_1 - \frac{1}{3}x_1^3 -x_2\right), \quad \dot{x}_2 = \frac{x_1}{\mu},
\end{aligned}
\end{equation}
for some $\mu > 0$.
Given an initial condition $x(0)$, let $\phi(t, x(0))$ be the solution to~\eqref{eq:vanderpol}. 
For $t_f > 0$, we use the approach from~\ref{subsec:LP_reachset} with the JAX interval analysis toolbox \immrax{}~\cite{harapanahalliImmraxParallelizableDifferentiable2024} to compute a polytope containing $\phi(t_f,x(0))$ for any initial condition inside a given box $x(0)\in[\ulx_0,\olx_0]$.
\end{example}

Example~\ref{ex:reach} requires the frequent solving of small LPs as in Algorithm~\ref{alg:control_filtering} (where each LP is automatically generated and contains dependent constraints). 
A plot of the reachable sets generated by our approach is given in Figure~\ref{fig:vdp_linprog}, and results are summarized in Table~\ref{table:intref_comp}. 
Here, \linrax{} excels, an order of magnitude faster than any other implementation. 
In fact, we are \emph{still} the fastest even when including JIT compile time, showing that the performance gains more than outweigh the oft-cited drawback of compilation times. 
This is due to the extra benefits of embedding the LP directly into the reachable set pipeline, allowing us to \kw{jit} the entire procedure.
The two gradient-based libraries perform especially poorly, with \cvxpy{} crashing before the test is complete and \jaxopt{} failing to converge to its default tolerance before the iteration limit for many programs, resulting in the slowest time and significantly worse final bound size.

\begin{table}
\centering
\begin{tabular}{|c|c|c|c|}
    \hline
    Method & \kw{jit} (sec)  & Time (sec) & Bound Size\\
    \hline \hline 
    \scipy{} & - & $37.80653 \pm 0.99753$ & $7.00845\cdot 10^{-2}$  \\ 
    \hline
    \gurobi{} & - & $44.66002 \pm 2.20993$ & $6.8551\cdot 10^{-2}$ \\ 
    \hline
    \cvxpy{} & - & Fail & - \\ 
    \hline
    \jaxopt{} & $\bf 4.1275$ & $56.64345 \pm 0.57630$ & $8.84713\cdot 10^{-2}$ \\ 
    \hline
    \bf{\linrax{}} & 6.25303 & $\bf 2.13317 \pm 0.054998$ & $6.8724 \cdot 10^{-2}$ \\ 
    \hline
\end{tabular}
\caption{%
    Comparison of LP solvers on the interval refinement problem Example~\ref{ex:reach} with $\mu=1$, $t_f = 0.628$.
    Sample size $N=10$.
    Bound size is quantified as the product of the widths of the interval bound for the reachable set; smaller sizes mean less conservatism. 
}
\vspace{-2mm}
\label{table:intref_comp}
\end{table}

\begin{figure}
\vspace{1mm}
    \centering
    \includesvg[width=\linewidth]{fig/vdp_linprog}
    \caption{Bounds on the reachable set of trajectories of the Van der Pol oscillator with $\mu=1$, $t_f = 2\pi$, $x_1(0) \in [0.9, 1.1]$, and $x_2(0) \in [-0.1, 0.1]$, computed with a LP-based interval refinement procedure. 
    Real Monte-Carlo sampled trajectories are shown in gray as a ``ground-truth'' comparison to these bounds. 
    Note how the bounds always contain every sampled trajectory, and remain relatively tight over a long time horizon. 
    }
    \label{fig:vdp_linprog}
\end{figure}


   
\section{Conclusions}

We have presented the first simplex-based LP solver compatible with the JAX ecosystem.
This work was motivated by two complementary trends: the increased use of optimization as a subroutine in control algorithms, and the growing popularity of JAX in controls applications. 
Stringent JAX requirements prevent direct implementation of standard simplex methods, particularly the handling of dependent constraints, a necessity in many algorithms such as the reachability-based safety control problem described in Section \ref{sec:safe_control}.
To overcome this limitation, we introduce a novel modification ensuring full compatibility with JAX requirements. 
We showed that when embedded into a JAX pipeline, \linrax{} outperforms state-of-the-art solvers, and demonstrated an application of automatic differentiation to solve safe robotics and controls problems.
An interesting possibility for future work would be to implement improvements to the simplex method, such as parallelization~\cite{huangfuParallelizingDualRevised2018}, or a more sophisticated pivoting rule.


\addtolength{\textheight}{-8cm}   

\bibliographystyle{ieeetr}
\bibliography{gould_ref}
\end{document}